\documentclass[a4paper,11pt]{article}
\pdfoutput=1
\usepackage{jcappub}
\usepackage{amssymb,amsmath,mathrsfs,enumerate}
\usepackage{graphicx,rotate,multicol}
\usepackage{float}

\usepackage{subfig}

\usepackage{slashed}
\usepackage{mathtools}
\usepackage{multirow}

\allowdisplaybreaks

\title{\boldmath Probing the Origin of Primordial Black Holes through Novel Gravitational Wave Spectrum}

\author{Indra Kumar Banerjee,}
\author{Ujjal Kumar Dey}
\affiliation{Department of Physical Sciences, Indian Institute of Science Education and Research Berhampur,\\Transit Campus, Government ITI, Berhampur 760010, Odisha, India}

\emailAdd{indrab@iiserbpr.ac.in}
\emailAdd{ujjal@iiserbpr.ac.in}

\abstract{In this article we investigate the cumulative stochastic gravitational wave spectra as a tool to gain insight on the creation mechanism of primordial black holes. We consider gravitational waves from the production mechanism of primordial black holes and from the gravitational interactions of those primordial black holes among themselves and other astrophysical black holes. We specifically focus on asynchronous bubble nucleation during a first order phase transition as the creation mechanism. We have used two benchmark phase transitions through which the primordial black holes and the primary gravitational wave spectra have been generated. We have considered binary systems and close hyperbolic interactions of primordial black holes with other primordial and astrophysical black holes as the source of the secondary part of the spectra. We have shown that this unique cumulative spectra have features which directly and indirectly depend on the specifics of the production mechanism.}

\begin{document}
\maketitle
\flushbottom


\section{Introduction}
\label{sec:intro}
The dark matter question has long been persecuting the all encompassing successes of the Standard Model (SM) of particle physics. In principle, dark matter could come in various shades, for example, it can be of particulate nature (e.g., weakly interacting massive particles, axions etc.) having widely distributed mass range or it can be some compact object of primordial origin, like primordial black hole (PBH).
%
%
Primordial back holes were first proposed in \cite{Zeldovich:1967lct}, and since then it has been the center of interest to the community. After the discovery of gravitational waves by the LIGO and Virgo collaborations (LVC)~\cite{LIGOScientific:2016aoc} almost a decade ago, the interest on PBHs was resurrected with new vigour. For long time many different mechanisms have been proposed regarding the origin of PBH, e.g. collapse from inhomogeneities during radiation-~\cite{Carr:1974nx,Grillo:1980rt} and matter-dominated era~\cite{Khlopov:1980mg, Khlopov:1985jw,Carr:1994ar,Chakraborty:2022mwu}, critical collapse~\cite{Niemeyer:1997mt, Niemeyer:1999ak}, collapse in single~\cite{Carr:1993aq, Bullock:1996at, Saito:2008em,Kawai:2021edk,Kawai:2021bye,Choudhury:2023hvf} and multi-field inflationary models~\cite{Randall:1995dj, Garcia-Bellido:2016dkw, Braglia:2020eai,Kawai:2022emp}, collapse of cosmic string loops~\cite{Hawking:1987bn, Borah:2023iqo}, collapse during a first order phase transition (FOPT)~\cite{Crawford:1982yz} to name a few. Recently, a number of novel mechanisms have been proposed related to the creation of PBHs during a FOPT. Some of these are model-dependent mechanisms, i.e. the particle content and the interactions of the model decides whether there will be any creation of PBH during a FOPT~\cite{Kawana:2021tde, Baker:2021nyl,Huang:2022him}, whereas the others are model-independent~\cite{Kawana:2022olo, Liu:2021svg,Gouttenoire:2023naa,Lewicki:2023ioy}. These model-independent mechanisms propose the creation of PBH through collapse of overdense regions in the early universe where these regions are formed due to the asynchronous nature of nucleation of true vacuum bubbles during a FOPT. 
Furthermore, many of these PBH forming mechanisms, i.e. scalar perturbation~\cite{Matarrese:1993zf, Matarrese:1996pp, Matarrese:1997ay}, inflation~\cite{Khlebnikov:1997di, Easther:2006vd, Easther:2007vj,Choudhury:2013woa}, cosmic strings~\cite{Vilenkin:1981bx, Vachaspati:1984gt, Hindmarsh:1994re}, FOPT~\cite{Witten:1984rs, Hogan:1986qda} can also be the sources of stochastic gravitational wave background (SGWB) in the Universe.
This SGWB and the properties of the PBH, i.e. their mass and abundance etc. depend on the mechanisms through which they were created. Apart from this, PBHs, if they exist, will interact gravitationally with themselves and other astrophysical black holes (ABH). In principle, these gravitational interactions can also create SGWB in the detectable range of the proposed GW detectors depending on the mass and the abundance of the PBHs. Therefore, we propose that along with the SGWB created directly by the creation mechanisms of PBH, (scalar perturbations, cosmic strings, FOPT, inflation, etc.) the SGWB created by those PBHs are also the consequence of those mechanisms. Therefore, the features of the cumulative SGWB is unique up to the mechanism that is responsible for the origin of PBH. To illustrate this further, the mass and abundance of the PBHs created during a FOPT depend on the specifics of the FOPT, i.e. the strength ($\alpha$), duration ($1/\beta$), energy content, critical temperature ($T_{\mathrm{cri}}$) etc. The SGWB produced due to the FOPT also depend on these same parameters. Therefore, the spectrum of the SGWB from the FOPT and the spectrum of the SGWB from the PBH interactions depend on the specifics of the FOPT directly and indirectly, respectively.
The literature on probing the PBHs utilising the features in the gravitational waves is quite large~\cite{Borah:2023iqo, Gehrman:2023esa, Franciolini:2022htd, Acuna:2023bkm, Gehrman:2022imk, Mandic:2016lcn, Chen:2018rzo, Sugiyama:2020roc, Garcia-Bellido:2021jlq, Cui:2021hlu, Papanikolaou:2022chm,Xie:2023cwi,Barman:2022pdo,Agashe:2022jgk}. However, an approach to study the cumulative effect of the GW background from the creation mechanism of PBH as well as the gravitational interactions of the PBHs is yet to be rigorously performed. In this article, we tried to fill that gap by investigating the sum of the SGWB due the FOPT and the SGWB due the interactions of PBH as prescription to test the claim that FOPTs are the origin of PBHs. In principle, this can be extended to all the other mechanisms which create both PBH and SGWB. Recently in Refs.~\cite{Gehrman:2023esa,Xie:2023cwi} it was proposed that SGWB spectra can be used to distinguish between creation mechanism of PBH; though the SGWB due to gravitational interactions of PBHs were not considered since for those mass ranges of PBH it would have been insignificant. Since we are considering gravitational interactions of the PBHs as one of the sources of SGWB, the mass of the PBHs are to be large enough for the SGWB to be significant. This led us to consider specific FOPTs, i.e. the FOPTs for which the transition temperature is below $\mathcal{O}(10\mathrm{~GeV})$ as examples which can generate PBHs with large masses, i.e. masses above $\mathcal{O}(10^{-5}M_{\odot})$. These comparatively low temperature FOPTs can be motivated from the dark sector phase transitions~\cite{Freese:2023fcr}. Owing to the larger mass, in this article we deal with PBHs for which Hawking evaporation \cite{Hawking:1974rv} does not play any major role. These PBHs are also created during the radiation dominated era, and therefore the PBH-PBH interactions in principle have started from a very early time in the universe, but the PBH-ABH interactions could have started only after the creation the of the first ABHs. We have taken these effects into account in our study as well. Furthermore, we also look into the dependence of the peak frequencies of the cumulative spectrum on the temperature at which the FOPT occurred. We also show the cumulative SGWB originating from FOPTs at different temperatures where some benchmark values of the relevant parameters have been used.
This article is organized as follows, in Sec. \ref{sec:fopt} we discuss FOPT as the common origin of SGWB and the PBHs. After discussing the mass and abundance of PBHs formed out of FOPT we go on to show the SGWB spectrum that can be originating from the FOPT. In Sec. \ref{sec:pbh_pbh} we discuss the SGWB spectrum originating from PBH-PBH interactions which can be of two main classes, namely the binary formations and closed hyperbolic encounters. Similar interactions can, in principle occur between PBH and ABH as well. The SGWB spectrum from out of PBH-ABH interactions is discussed in Sec. \ref{sec:pbh_abh}. We present our results in Sec. \ref{sec:results} and finally summarize and conclude in Sec. \ref{sec:concl}.

\section{FOPT as the Origin of the PBH and GW}
\label{sec:fopt}
Many different processes have been considered for the creation of PBHs with mass and abundance ranging throughout the parameter space. In this article we consider the creation of PBH from the collapse of overdense region during a first order phase transition. Here we consider the mechanisms proposed in~\cite{Kawana:2022olo,Liu:2021svg,Gouttenoire:2023naa,Lewicki:2023ioy}.

During its evolution as the temperature decreased, the universe might have gone through several phase transitions. These phase transitions are characterized by the potential of the order parameter that drives the transition. These phase transitions could be of first or second order in nature depending on the shape of the potential. If it has a barrier between the true vacuum (newly generated global minima) and the false vacuum (pre-existing yet local minima), then it is a first order phase transition, otherwise, it is a second order. The temperature, below which the pre-existing minima becomes a local minima and a new global minima is created, is called the critical temperature, $T_{\mathrm{cri}}$. A first order phase transition, which we are interested in here, occurs through the nucleation of true vacuum bubbles which then expand releasing huge amount of energy to the bubble walls and the surrounding plasma. The rate of the nucleation of the true vacuum is approximately given around an appropriate time $t_0$ by~\cite{Coleman:1977py},
\begin{align}
\Gamma\approx\Gamma_0 e^{\beta(t-t_0)}
\label{gamma_def}
\end{align}
where, 
\begin{align}
\beta=-\dfrac{dS(t)}{dt}\Big\vert_{t=t_0}
\label{beta_def}
\end{align}
and $S(t)$ is the bounce action of the four or three dimensional instanton solution depending on the temperature of the transition~\cite{Laine:2016hma}.
Before the universe reaches the critical temperature, the average fraction of false vacuum is unity. But as the phase transition progresses, the average fraction of false vacuum reduces depending of the specific dynamics of the phase transition. Since the nucleation of true vacuum bubble is a probabilistic process, in some Hubble volume, the creation of a true vacuum bubble might be delayed. The idea is that as the time passes and the universe expands, the radiation and the wall energy density in the false vacuum region falls as $\propto a^{-4}(t)$ though the vacuum energy density remains constant. Soon after, the vacuum energy in the false vacuum region starts decaying into the radiation and wall energy density. In the true vacuum region, since the vacuum energy density is much less than that of the false vacuum region, the ratio of energy density inside and outside the false vacuum region keeps increasing and reaches a maximum value when the entire vacuum energy inside the false vacuum region has been decayed. Once it reaches a critical value of $1.45$, the false vacuum region collapses to form a PBH~\cite{Musco:2004ak}. For a more in-depth explanation, see~\cite{Kawana:2022olo,Liu:2021svg,Gouttenoire:2023naa,Lewicki:2023ioy}.

Now we briefly discuss the mass and the abundance of the PBHs arising due to different cases of FOPT through the method mentioned above. In this work we consider two benchmark FOPTs, the parameters of which are provided in Tab.~\ref{table_FOPT}. Transitions around such kind of temperatures can arise in various BSM scenarios~\cite{Freese:2023fcr,Ellis:2022lft}. Moreover, these FOPTs would generate masses for the PBH which are large enough for our purposes while maintaining the various abundance constraints depicted in Fig.~\ref{mass_abun}.
\begin{table}[H]
\centering
\begin{tabular}{|l|l|l|l|}
\hline
\multicolumn{1}{|c|}{Parameters} & \multicolumn{1}{c|}{I} & \multicolumn{1}{c|}{II} \\ \hline \hline
$T$ (in GeV) & 0.06  & 45  \\ \hline
$\alpha$    & 1                   & 1.5 \\ \hline
$\beta/H$   & 3.5                  & 2.5 \\ \hline
$\kappa$   & 0.5                  & 1 \\ \hline
\end{tabular}
\caption{The relevant parameters describing the different FOPTs.}
\label{table_FOPT}
\end{table}
Here $\kappa$ is the fraction of vacuum energy transferred into the bubble walls. Now we will use these values to obtain the mass and the abundance of the PBHs and the gravitational waves due to collision of the bubble walls.
\subsection{Creation of PBHs: Mass and Abundance}
\label{cre_pbh}

%
The formation mechanism of PBHs that we are following~\cite{Kawana:2022olo, Liu:2021svg, Gouttenoire:2023naa, Lewicki:2023ioy}, is based on the collapse of overdense regions which are a consequence of delayed nucleation of true vacuum bubbles. However, there are slight differences in the specifics of the creation mechanisms of~\cite{Kawana:2022olo, Liu:2021svg, Gouttenoire:2023naa, Lewicki:2023ioy}, i.e. the contribution of the nucleation in the past light-cone which leads to slightly different abundance of PBHs for the same parameter values. It is also to be kept in mind that the mass of the PBHs are dependent on the nucleation temperature and therefore is of the same order for all the cases. To quantify the PBH parameters we note that the mass of the PBH formed due to the above-mentioned process can approximately be given as,
\begin{align}
M(t)=\gamma\frac{4\pi}{3}\bar{\rho}(t)H^{-3}(t),
\label{pbh_mass}
\end{align}
where, $H^2(t) = \bar{\rho}(t)/3M^2_{\text{P}},$
with $\bar{\rho}$ is the average energy density and $M_{\text{P}}$ is the Planck mass. Here $\gamma$ ($\le 1$) is a numerical factor which depends on the specifics of the gravitational collapse. 
Now, we assume that most of the PBHs which are formed during a FOPT have a mass, $M_{\mathrm{PBH}}=M(t_{1.45})$, where $t_{1.45}$ is defined by,
\begin{align}
\frac{\rho_{\mathrm{inside}}(t_{1.45})}{\rho_{\mathrm{outside}}(t_{1.45})}=1.45.
\end{align}
Using this prescription, the masses of PBHs formed during FOPT I and FOPT II are $40M_{\odot}$ and $3\times 10^{-5}M_{\odot}$ respectively.
%

%

\begin{figure}[H]
\centering
\includegraphics[scale=0.3]{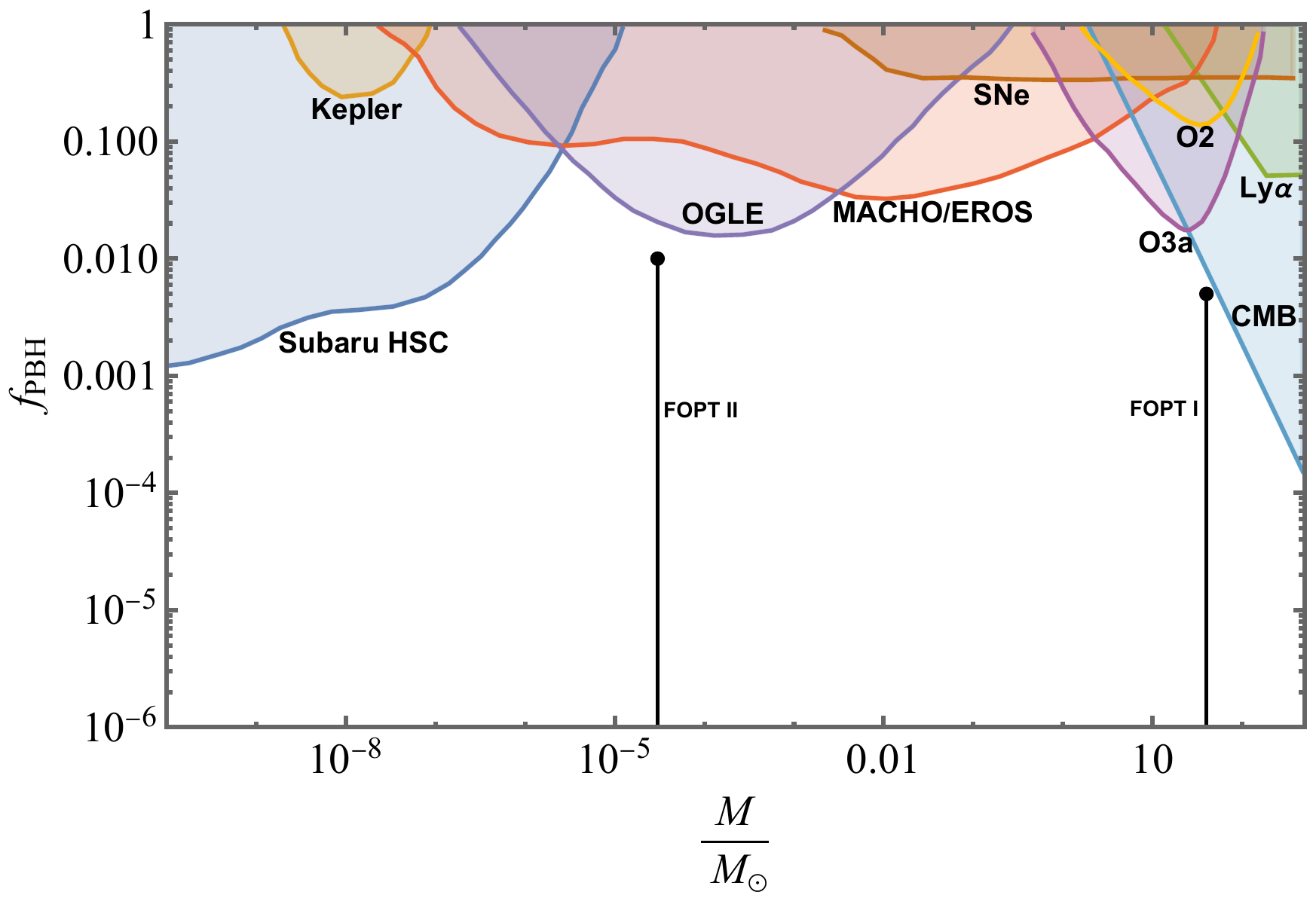}
\caption{Abundances for different PBH masses. The constraints from Subaru HSC~\cite{Niikura:2017zjd}, Kepler~\cite{Griest:2013esa}, OGLE~\cite{Niikura:2019kqi}, MACHO/EROS~\cite{Macho:2000nvd,EROS-2:2006ryy}, SNe~\cite{Zumalacarregui:2017qqd}, Ly-$\alpha$~\cite{Murgia:2019duy}, CMB~\cite{Poulin:2017bwe} and the non-observation of SGWB from the second observing run (O2) and the first half of the third observing run (O3a) of LVC~\cite{Hutsi:2020sol} have also been shown.}
\label{mass_abun}
\end{figure}

Slightly different PBH abundances can be generated from similar parameter ranges~\cite{Kawana:2022olo, Liu:2021svg, Gouttenoire:2023naa, Lewicki:2023ioy}. Therefore, remaining agnostic about the finer details of the mechanisms, we take two benchmark values for the PBH abundances, i.e. for the PBHs created during to FOPT I, the abundace today is $0.005$ whereas for the PBHs created during the FOPT II, today's abundance is $0.01$.
The mass of the PBHs and their abundances along with the relevant constraints\footnote{Since the LVC constraints from the merger rates depend on the nature of the events, i.e. whether some or all of the events are of primordial nature~\cite{Hutsi:2020sol}, we have not included those.} have been depicted in Fig.~\ref{mass_abun}.

\subsection{GW Background from FOPT}
\label{gw_fopt}
\begin{figure}[H]
\centering
\includegraphics[scale=0.4]{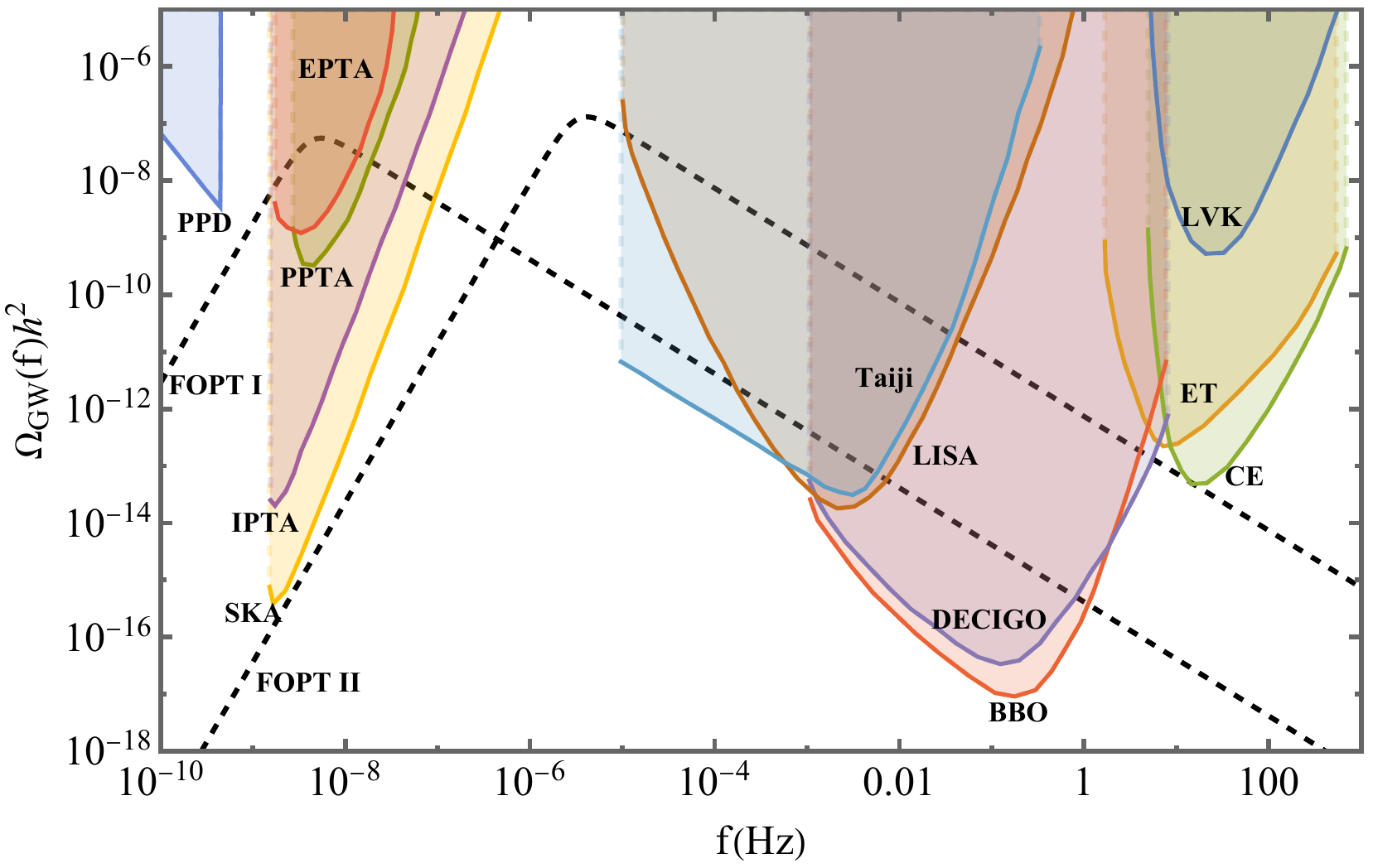}
\caption{The SGWB spectrum for the two first order phase transitions considered in this paper. The sensitivity curves for SKA~\cite{Carilli:2004nx}, IPTA~\cite{Hobbs:2009yy}, EPTA~\cite{Lentati:2015qwp}, PPTA~\cite{Shannon:2015ect}, Taiji~\cite{Ruan:2018tsw}, LISA~\cite{LISA:2017pwj}, DECIGO~\cite{Kawamura:2011zz}, BBO~\cite{Phinney:2004bbo}, CE~\cite{Reitze:2019iox}, ET~\cite{Punturo:2010zz}, LIGO, Virgo and KAGRA (LVK)~\cite{Somiya:2011np,LIGOScientific:2014pky} have been shown. The constraints from pulsar parameter drifts (PPD)~\cite{DeRocco:2023qae} have also been shown.}
\label{fopt_sgwb}
\end{figure}

Now we give a short description of the SGWB spectrum that can be generated from the FOPTs. The expression for the SGWB spectrum due to FOPT can be expressed as~\cite{Kosowsky:1992vn},
\begin{align}
\Omega_{\mathrm{GW}}(f)&=\dfrac{1}{\rho_c}\dfrac{d\rho_{\mathrm{GW}}}{d\ln f} \nonumber \\
&=1.67\times 10^{-5}\left(\dfrac{H}{\beta}\right)^2 \left(\dfrac{\kappa\alpha}{1+\alpha}\right)^2 \dfrac{0.11v_w^2}{0.42+v_w^3}\left(\dfrac{100}{g_{*}}\right)^{1/3} \dfrac{3.8(f/f_p)^{2.8}}{1+2.8(f/f_p)^{3.8}},
\label{FOPT_GW}
\end{align}
where,
\begin{align}
f_p=\dfrac{0.62}{1.8-0.1v_w+v_w^2}\left(\dfrac{\beta}{H}\right)\dfrac{T}{100\mathrm{~GeV}}\left(\dfrac{g_{*}}{100}\right)^{1/6}\times 1.65\times 10^{-5} \mathrm{~Hz},
\label{FOPT_fp}
\end{align}
is the peak frequency and $v_w$ is the bubble wall velocity, the value of which has been taken to be unity. It is worth-mentioning that in the present case we only consider the contributions from collision of bubble walls as they contribute the most to the SGWB spectrum. We also consider that the phase transition occurs fast enough and therefore all the parameters have been evaluated as $T_{\mathrm{cri}}$.
Now using the information in Tab. \ref{table_FOPT} in Eqs.~\eqref{FOPT_GW} and \eqref{FOPT_fp} we get the gravitational wave spectrum which has been presented in Fig. \ref{fopt_sgwb}.
\section{PBH-PBH Interactions}
\label{sec:pbh_pbh}
The PBHs after being formed during the FOPT are free to interact among themselves. Depending on the initial conditions, e.g relative velocity and the scattering cross sections, etc., two PBHs can either form a bound binary system or there could be a single scattering event~\cite{Kocsis:2006hq,OLeary:2008myb,Capozziello:2008ra,DeVittori:2012da,Garcia-Bellido:2017knh,Garcia-Bellido:2021jlq}. For binary black holes (BBH) there could be three cases where different types of GWs can be produced -- (i) individual mergers which create short-lived transient signals, (ii) binaries which are in inspiraling phase create GW signals that last longer than the observational period, and (iii) binaries which are unresolvable and create SGWB~\cite{Pujolas:2021yaw}. 
Similarly, individual scattering events or close hyperbolic encounters (CHE) of two PBHs can also either create transient signals or they can create SGWB. In this article, our main focus is on the SGWB spectrum created by these events.
As opposed to the previous section, where the source of the SGWB spectrum was considered to be the FOPT, here, we consider many individual unresolvable events. These individual events can be considered to be point-like. In case of a point source, the SGWB spectrum is expressed as,
\begin{align}
\Omega_{\mathrm{GW}}(f)&=\dfrac{1}{\rho_c}\dfrac{d\rho_{\mathrm{GW}}}{d\ln f} \nonumber \\
&=\dfrac{1}{\rho_c}\int_0^{\infty} dz \dfrac{N(z)}{1+z}\dfrac{dE_{\mathrm{GW}}}{d\ln f_s},
\label{sgwbspectgen}
\end{align}
where $\rho_c$ is the critical density of the universe, $f_s=(1+z)f$ is the GW frequency in the source frame and $z$ is the redshift. Here $N(z)$ is the number density of GW events, at redshift $z$ and it is given by,
\begin{align}
N(z)=\dfrac{\tau(z)}{(1+z)H(z)},
\label{numdensgwevents}
\end{align}  
where $H(z)$ is the Hubble expansion coefficient and $\tau(z)$ is the number of events per unit time per unit comoving volume and it is expressed in units of $\mathrm{yr}^{-1}\mathrm{Gpc}^{-3}$. Now, in case of different mass spectrum for two different progenitor  masses this is can be given as,
\begin{align}
\tau(z)=\int \int \dfrac{dm_1}{m_1}\dfrac{dm_2}{m_2}\dfrac{d\tau}{d\ln m_1 d\ln m_2}.
\label{mergrate}
\end{align}
Here we consider the number of event to be independent of the redshift. The quantities $dE_{\mathrm{GW}}/d\ln f_s$ and $\tau(z)$ depend on the event type (BBH or CHE). In the next part of the article, we will focus specifically on these different type of events and generate the relevant GW spectrum.

\subsection{Binary PBH systems}
\label{pbh_bbh}
If the PBHs are clustered in dense halos, then their number density can be expressed as a combination of their mass function, the critical density of the universe, their masses and the local density contrast. In case of PBH binary systems, the event rate per logarithmic mass interval takes the form~\cite{Garcia-Bellido:2021jlq},
\begin{align}
\dfrac{d\tau_{\mathrm{BBH}}}{d\ln m_1 d\ln m_2}=14.8\mathrm{~yr}^{-1}\mathrm{Gpc}^{-3}
f(m_1) f(m_2) \dfrac{M^{10/7}}{(m_1 m_2)^{5/7}} \left(\dfrac{\Omega_{\mathrm{DM}}}{0.25}\right)^2 \left(\dfrac{\delta_{\mathrm{loc}}}{10^8}\right) \left(\dfrac{v_0}{10\mathrm{~km/s}}\right)^{-11/7} ,
\label{mergrate_bbh}
\end{align}
where $\Omega_{\mathrm{DM}}$ is the density parameter of dark matter in the universe, $\delta_{\mathrm{loc}}$ is the local density contrast which can be considered to be $\mathcal{O}(10^8)$~\cite{Clesse:2016ajp}, $v_0$ is the initial relative velocity between the two PBHs participating in the binary formation, $M=(m_1 + m_2)$ is the total mass where $m_1$ and $m_2$ are the masses of the individual masses of the PBHs and $f(m_1)$ and $f(m_2)$ are the logarithmic mass function of the PBHs of masses $m_1$ and $m_2$ respectively such that $\int f(m) dm/m=f_{\mathrm{PBH}}$ where $f_{\mathrm{PBH}}$ is the fraction of dark mater which is constituted of PBHs. It is to be noted that in this article we have considered the two different dimensionless Hubble parameters to be  $h_{70}=1$ and $h=0.7$.
In this case since we consider the collapse of overdense region during a FOPT as the creation mechanism, the logarithmic mass function of the PBH of mass $m'$ is $\propto \delta(m-m')$, i.e. we assume that all the PBHs created due to this mechanism are of the same mass. 

\begin{figure}[H]
\centering
\includegraphics[scale=0.38]{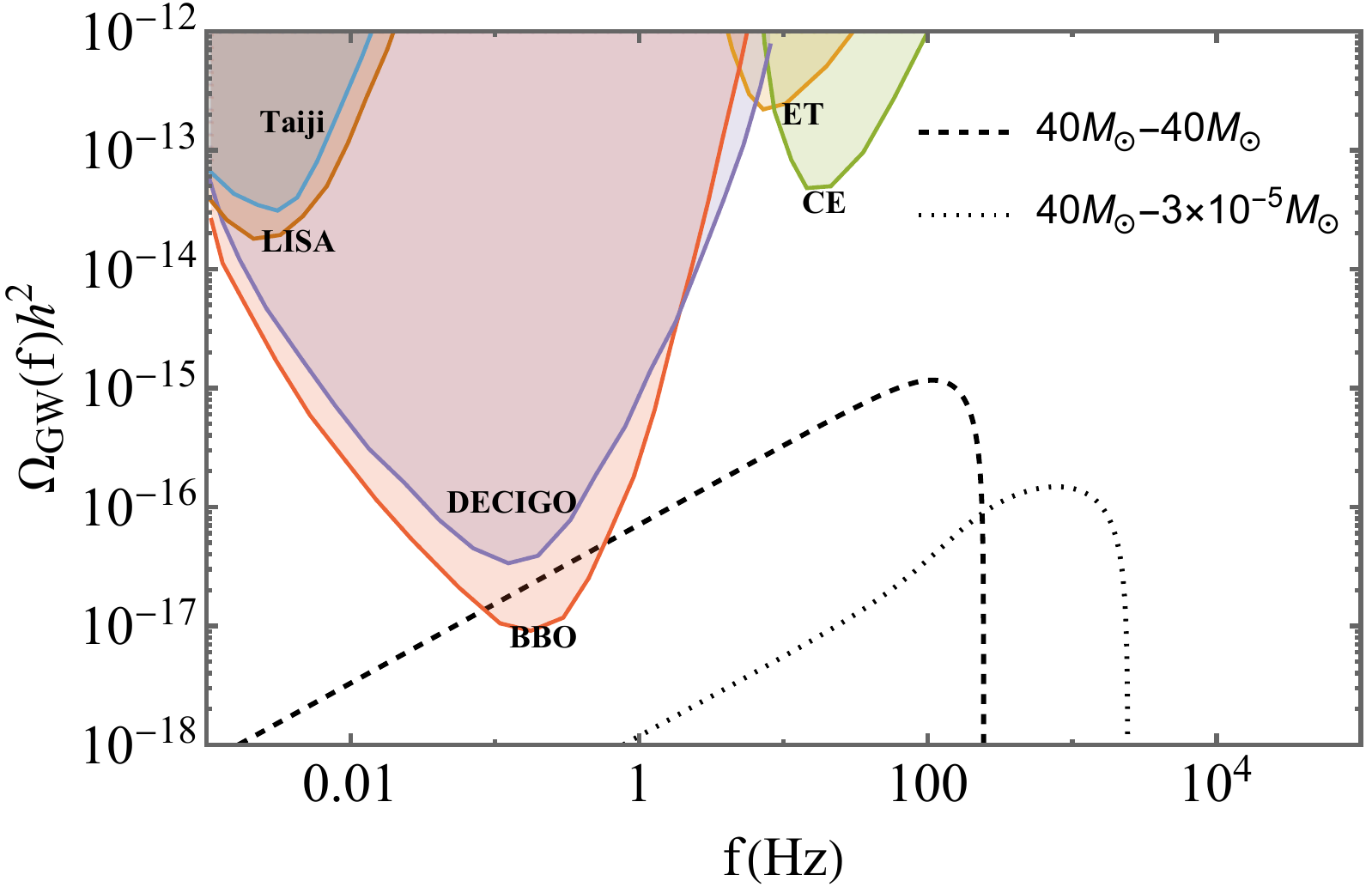}
\caption{The SGWB spectrum for PBH binaries of mass combinations $40M_{\odot}-40M_{\odot}$ and $40M_{\odot}-3\times 10^{-5}M_{\odot}$. The relevant sensitivity curves have been shown.}
\label{pbh_bbh_final}
\end{figure}

Now we consider the energy emitted as GW per frequency interval, i.e. $dE_{\mathrm{GW}}/df_s$ in case of a BBH. For this purpose, we use the waveform for non-spinning binaries~\cite{Ajith:2007kx},
\begin{align}
\dfrac{dE_{\mathrm{GW}}}{df_s}=\dfrac{(G\pi)^{2/3}M_c^{5/3}}{3}
\begin{dcases}
        f_s^{-1/3}, & f_s \leq f_{\mathrm{merg}} \\
        f_{\mathrm{merg}}^{-1}f_{s}^{2/3}, & f_{\mathrm{merg}}\leq x\leq f_{\mathrm{ring}} \\
        f_{\mathrm{merg}}^{-1}f_{\mathrm{ring}}^{-4/3}f_s^2\left[\left(\dfrac{f_s-f_{\mathrm{ring}}}{\sigma /2}\right)^2 +1\right]^{-2}, & f_{\mathrm{ring}}\leq f_s \leq f_{\mathrm{cut}} \\
        0, & f_s > f_{\mathrm{cut}}\\
    \end{dcases}
\label{energy_bbh}
\end{align}
where, $M_c=(m_1 m_2)^{3/5}/(m_1 +m_2)^{1/5}$ is the chirp mass. A binary contributes to the GW spectrum in three different phases, the inspiral, the merger and the ringdown. The inspiral and the merger phases are separated by $f_{\mathrm{merg}}$ and the merger and the ringdown phases are separated by $f_{\mathrm{ring}}$. Here, $f_{\mathrm{cut}}$ signifies the frequency where the binary stops producing GWs and $\sigma$ signifies the width of the transition from merger to ringdown stage. Here, these four parameters can be expressed as,
\begin{align}
\epsilon_i=\dfrac{a_i\eta^2 +b_i\eta +c_i}{\pi M}\times \dfrac{c^3}{G},
\end{align}
where $\epsilon_1$, $\epsilon_2$, $\epsilon_3$, and $\epsilon_4$ signifies $f_{\mathrm{merg}}$, $f_{\mathrm{ring}}$, $\sigma$, and $f_{\mathrm{cut}}$, respectively. It is to be noted that  $\eta = (m_1 m_2)/(m_1 + m_2)^2$ is the symmetric mass ratio. The values of $a_i,~b_i$ and $c_i$~\cite{Ajith:2007kx} have been mentioned in Tab.~\ref{abc_table}.
\begin{table}[H]
\centering
\begin{tabular}{|l|l|l|l|}
\hline
\multicolumn{1}{|c|}{$i$} & \multicolumn{1}{c|}{$a_i$} & \multicolumn{1}{c|}{$b_i$} & \multicolumn{1}{c|}{$c_i$} \\ \hline \hline
1                         & $2.97\times 10^{-1}$    & $4.48\times 10^{-2}$    & $9.56\times 10^{-2}$    \\ \hline
2                         & $5.94\times 10^{-1}$    & $8.98\times 10^{-2}$    & $1.91\times 10^{-1}$    \\ \hline
3                         & $5.08\times 10^{-1}$    & $7.75\times 10^{-2}$    & $2.24\times 10^{-2}$    \\ \hline
4                         & $8.48\times 10^{-1}$    & $1.28\times 10^{-1}$    & $2.73\times 10^{-1}$    \\ \hline
\end{tabular}
\caption{The values to determine the relevant parameters for the GW spectrum of BBH.}
\label{abc_table}
\end{table}
It is to be noted that, in this case we have used non-spinning PBHs. If the PBHs were spinning, then the contributions from the BBH would have been different, which we leave for future work. Now we use Eqs.~\eqref{energy_bbh} and \eqref{mergrate_bbh} in Eq.~\eqref{sgwbspectgen} to obtain SGWB spectrum for the different combinations of PBH masses. For binaries consisting of two $3\times 10^{-5}M_{\odot}$ PBHs, the spectrum is insignificant, therefore we have not shown it. The other two spectrum have been shown in Fig. \ref{pbh_bbh_final}. It can be seen that for the $40M_{\odot}$ binaries, the SGWB spectrum is well within the reach of BBO. But for the other two cases, the SGWB spectrum is not within the sensitivity of any proposed GW detector.
It is also to be noted here that though inspiral, merger and ringdown, all three of them contribute to the GW spectrum, almost all of the contributions come from the inspiral phase, i.e. $f < f_{\mathrm{merg}}$. It can also be observed from the plots that the peak frequency for SGWB due to BBH is approximately,
\begin{align}
f_{\mathrm{peak}}\approx 0.7\times f_{\mathrm{merg}}.
\label{fpeak_bbh}
\end{align}
We use this above relation in Sec.~\ref{sec:results} in order to determine the dependence of this peak frequency on the temperature at which the FOPT occurred.

\subsection{Close Hyperbolic Encounters of PBH}
\label{pbh_che}
\begin{figure}[H]
\centering
\includegraphics[scale=0.34]{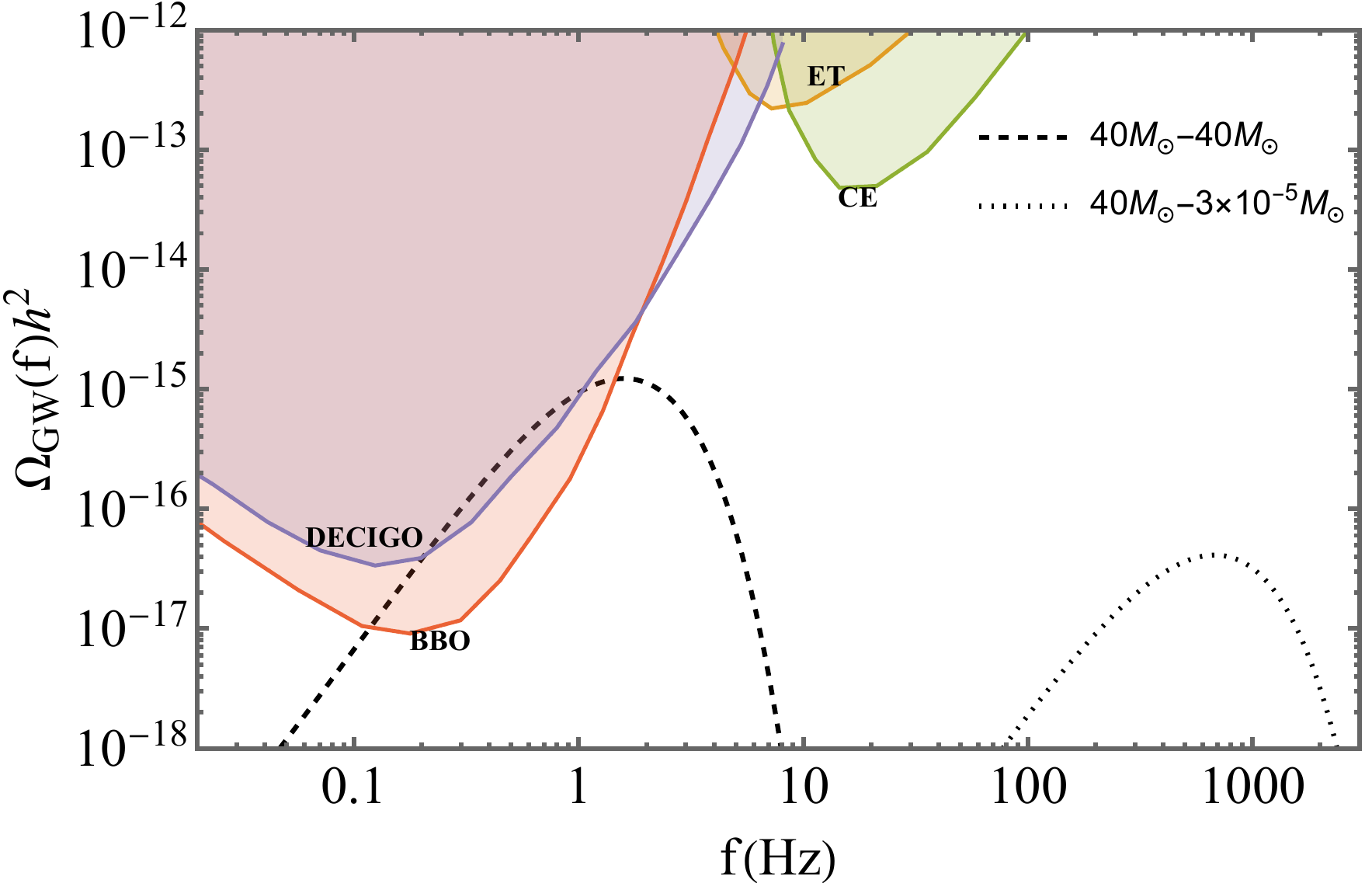}
\caption{The SGWB spectrum for close hyperbolic encounters of PBHs of mass combinations $40M_{\odot}-40M_{\odot}$ and $40M_{\odot}-3\times 10^{-5}M_{\odot}$. Here, for the first combination $a=7\times 10^{4}\mathrm{~AU}$ and $y=3.5\times 10^{-5}$ and for the second combination $a=3.5\times 10^{5}\mathrm{~AU}$ and $y=2\times 10^{-6}$. The relevant sensitivity curves have been shown.}
\label{pbh_che_final}
\end{figure}

In addition to the BBH scenario, as discussed previously, the PBHs can take part in close hyperbolic encounters (CHE). Such encounters can also give rise to specific type of GW spectrum. For the CHE, the event rate can be expressed as~\cite{Garcia-Bellido:2021jlq},
\begin{align}
\dfrac{\tau_{\mathrm{CHE}}}{d\ln m_1d\ln m_2}=25.4\times 10^{-8} \left(\dfrac{\Omega_{\mathrm{DM}}}{0.25}\right)^2 \left(\dfrac{\delta_{\mathrm{loc}}}{10^8}\right)f(m_1)f(m_2) \dfrac{M^2}{m_1 m_2} \dfrac{e^2-1}{(v_0/c)^3},
\end{align} 
where $e$ is eccentricity of the path taken by the PBH. Now the energy per logarithmic frequency interval can be expressed as~\cite{DeVittori:2012da,Garcia-Bellido:2017qal},
\begin{align}
\dfrac{dE_{\mathrm{GW}}}{d\ln f_s}=\dfrac{4\pi}{45}\dfrac{G^3 m_1^2 m_2^2}{a^2 c^5 \nu_0}\nu^5 F_e(\nu),
\end{align}
where $a$ is the semi-major axis, $\nu=2\pi\nu_0 f_s$ and $\nu_0=a^3/GM$.
We use the prescription given in Ref.~\cite{Garcia-Bellido:2021jlq} to obtain the expression for peak frequency and the SGWB spectrum.
The peak frequency for the SGWB due to the CHEs of the PBHs takes the form~\cite{Garcia-Bellido:2021jlq},
\begin{align}
f_{\mathrm{peak}}\sim 43\mathrm{~Hz} \left(\dfrac{y}{0.01}\right)^{-3}\left(\dfrac{M}{200~M_{\odot}}\right)^{1/2}\left(\dfrac{a}{0.1\mathrm{~AU}}\right)^{-3/2},
\label{fpeak_che}
\end{align}
and the associated SGWB spectrum takes the form,
\begin{align}
\Omega_{\mathrm{GW}}(f) = 9.81 &\times 10^{-13}   \left(\dfrac{\Omega_{\mathrm{DM}}}{0.25}\right)^2 \left(\dfrac{\delta_{\mathrm{loc}}}{10^8}\right) \left(\dfrac{a}{0.1\mathrm{~AU}}\right) \left(\dfrac{f}{10\mathrm{~Hz}}\right)^2 \left(\dfrac{y}{0.01}\right)\nonumber\\
&\times \int \dfrac{dm_1}{100~M_{\odot}}\dfrac{dm_2}{100~M_{\odot}} f(m_1) f(m_2) e^{-2 x_0 \zeta(y)} \tilde{I}(y,x_0),
\end{align}
here we have used density parameter of matter in the universe, $\Omega_{\mathrm{M}}=0.3$. The quantities $x_0$, $\zeta(y)$, $\tilde{I}(y,x_0)$ are defined in Ref.~\cite{Garcia-Bellido:2021jlq}. 
Using these expressions we have obtained the SGWB spectrum for CHEs of the different combinations of the PBH masses. As expected, similar to the case of BBHs, here also, the CHEs between two PBHs of mass $3\times 10^{-5}M_{\odot}$ are insignificant and we have not shown it. The SGWB spectrum for the other two mass combinations have been shown in Fig. \ref{pbh_che_final}. It is to be noted that the spectrum for the CHEs between two $40M_{\odot}$ PBHs are within the sensitivity curves of BBO and DECIGO, but the spectrum due to the CHEs between a $40M_{\odot}$ and a $3\times 10^{-5}M_{\odot}$ is not within the sensitivity curves of any of the futuristic or current GW detection experiments.
%

\section{PBH-ABH Interactions}
\label{sec:pbh_abh}
If PBHs exist, then there should necessarily be the PBH-ABH interactions~\cite{Cui:2021hlu,Kritos:2020wcl} which should consist of both bound PBH-ABH systems and also single scattering events. In the subsequent section of the article, we take both of them into account and consider the respective GW spectrum.
\subsection{PBH-ABH Binary Systems}
\label{pbh_abh_bbh}
\begin{figure}[H]
\centering
\includegraphics[scale=0.32]{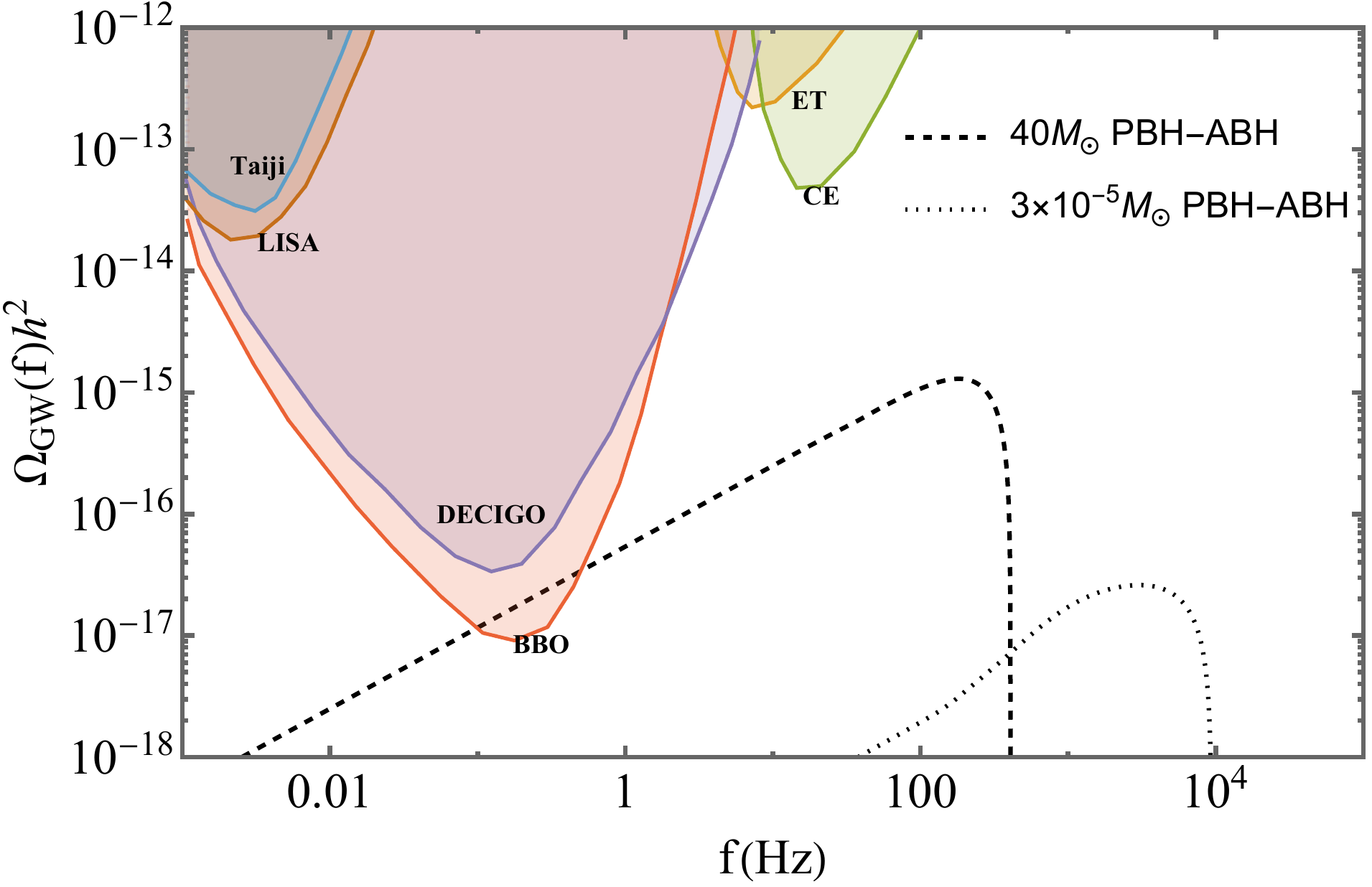}
\caption{The SGWB spectrum for PBH-ABH binaries for PBH mass $40~M_{\odot}$ and $3\times 10^{-5}M_{\odot}$. The relevant sensitivity curves have been shown.}
\label{pbhabh_bbh_final}
\end{figure}

First we take up the case of PBH-ABH binaries and how they contribute to the SGWB spectrum. In order to obtain the SGWB spectrum, we use the approach similar to Sec.~\ref{sec:pbh_pbh}. The averaged number of ABH per solar mass of stellar objects turns out to be $\sim 2\times 10^{-3}$~\cite{Cui:2021hlu}.
The merger rate for PBH-ABH binaries per logarithmic mass interval can be expressed as~\cite{Clesse:2016ajp},
\begin{align}
\dfrac{d\tau_{\mathrm{BBH}}}{d\ln m_1 d\ln m_2}=\dfrac{1}{\delta_{\mathrm{loc}}}\sigma_{\mathrm{BBH}}v_{\mathrm{BH}}n_{A}(m_A)n_{P}(m_{P}),
\label{mergrate_pbhabh}
\end{align}
where $n_A$ is the number density of the ABHs, $n_{P}$ is the number density of the PBHs, and $v_{\mathrm{BH}} = v_0 /\sqrt{2}$, where $v_0$ is the initial relative velocity between the PBH and the ABH. Here, $\sigma_{\mathrm{BBH}}$ is the cross section of the interactions between ABH and PBH in case of the formation of a binary which is given by~\cite{Mouri:2002mc},
\begin{align}
\sigma_{\mathrm{BBH}}=\pi\left(\dfrac{340\pi}{3}\right)^{2/7}\dfrac{G^2M^{10/7}(m_1m_2)^{2/7}}{c^{10/7}v_0^{18/7}}.
\label{cros_sec}
\end{align}
For our numerical analysis, we assume that ABH mass have ranges from $5M_{\odot}$ to a few tens of solar masses~\cite{Bailyn:1997xt,Farr:2010tu}. In this case, irrespective of when the PBH formation took place, the formation of binaries can only start after the creation of the first ABH. Therefore, unlike Eq.~\eqref{sgwbspectgen}, one can not take the upper-limit of the redshift integration to be infinite. For the purposes of this work, we have considered the upper limit of the redshift for the PBH-ABH interactions to be $5$~\cite{Cui:2021hlu}. Hence, Eq.~\eqref{sgwbspectgen} takes a modified form,
\begin{align}
\Omega_{\mathrm{GW}}(f)
&=\dfrac{1}{\rho_c}\int_0^{5} dz \dfrac{N(z)}{1+z}\dfrac{dE_{\mathrm{GW}}}{d\ln f_s}.
\label{sgwbspectgenpbhabh}
\end{align}
Now we use Eqs.~\eqref{mergrate_pbhabh} and \eqref{cros_sec} in Eqs.~\eqref{numdensgwevents} and \eqref{sgwbspectgenpbhabh} to obtain the SGWB spectrum for PBH-ABH binaries. In Fig.~\ref{pbhabh_bbh_final} we have shown the SGWB spectrum for PBH-ABH binaries with PBH mass $40M_{\odot}$ and $3\times 10^{-5}M_{\odot}$. Again as expected, it can be seen that for $40M_{\odot}$ PBHs, the SGWB spectrum due to the PBH-ABH binaries are within the reach of BBO, whereas for $3\times 10^{-5}M_{\odot}$ PBHs, the PBH-ABH binaries create SGWB which is too faint to be detected by any of the proposed detectors.
%

\subsection{PBH-ABH Close Hyperbolic Encounters}
\label{pbh_abh_bbh}
\begin{figure}[H]
\centering
\includegraphics[scale=0.4]{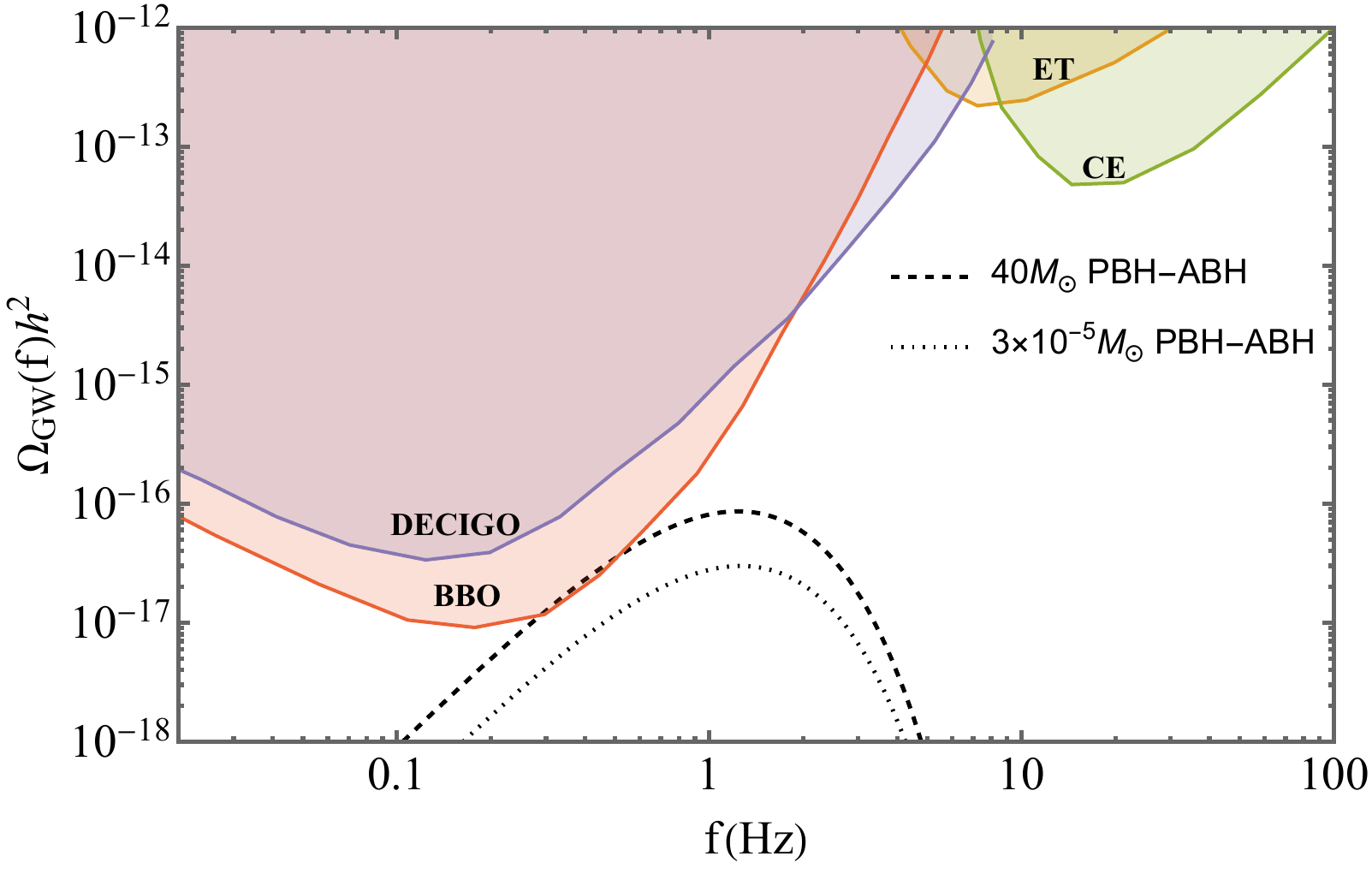}
\caption{The SGWB spectrum for PBH-ABH close hyperbolic encounters for PBH mass $40M_{\odot}$ and $3\times 10^{-5}M_{\odot}$. Here, for $40M_{\odot}$ PBHs $a=7\times 10^4\mathrm{~AU}$ and $y=3.5\times 10^{-5}$ and for $3\times 10^{-5}M_{\odot}$ PBHs $a=3.5\times 10^5\mathrm{~AU}$ and $y=2\times 10^{-6}$. The relevant sensitivity curves have been shown.}
\label{pbhabh_che_final}
\end{figure}

For completeness, we also consider the PBH-ABH close hyperbolic encounters as a source of the SGWB. In order to obtain the SGWB spectrum due to the PBH-ABH CHE, we follow a prescription similar to that of PBH-PBH CHE. In this case, the main difference is the number density of the ABHs which has been explained in the previous sub-section. Taking that into account, we have generated the SGWB spectrum for PBH-ABH CHE for both the PBH masses considered in this article and they have been shown in Fig. \ref{pbhabh_che_final}. It is to be noted that for the CHEs of $40M_{\odot}$ PBHs with the ABHs, the SGWB is marginally within the sensitivity curve of BBO, but in case of $3\times 10^{-5}M_{\odot}$ PBHs, the resultant SGWB is well beyond the reach of any of the proposed GW detectors. This is again consistent with the basic expectation that more massive encounters will lead to GW with larger peak frequencies.

\section{Results}
\label{sec:results}
In this article, we consider a situation where PBHs are generated due to the cosmological FOPTs, therefore along with the SGWB signal generated directly through the bubble wall collisions during the FOPT, the SGWB signals due the interactions of these PBHs among themselves and with other ABHs are also indirectly a consequence of the FOPT. So, we look at the sum of all these SGWB signals, which we call the cumulative background spectrum. 
We present our results in two main sections. First we elaborate on the cumulative stochastic background due to a first order phase transition and then we focus on the peak frequencies of the different components of the cumulative background and their dependence on the transition temperature. Along with this, we also look at the dependence of the cumulative background on the temperature of the FOPT.
\subsection{Cumulative Background}
\label{cum_bkg}
As we have seen in Sec. \ref{sec:pbh_pbh}, the SGWB originating from the PBH binary systems and the single scattering events are only significant and detectable by the future GW detectors for the cases where two $40 M_{\odot}$ interacts with each other or where $40 M_{\odot}$ PBH interacts with ABH. Therefore, in this part, we just focus on that result, i.e. the SGWB from the FOPT at $60\mathrm{~MeV}$ and the SGWB from the interactions of the PBHs created during this FOPT.

\begin{figure}[t]
\centering
\includegraphics[scale=0.5]{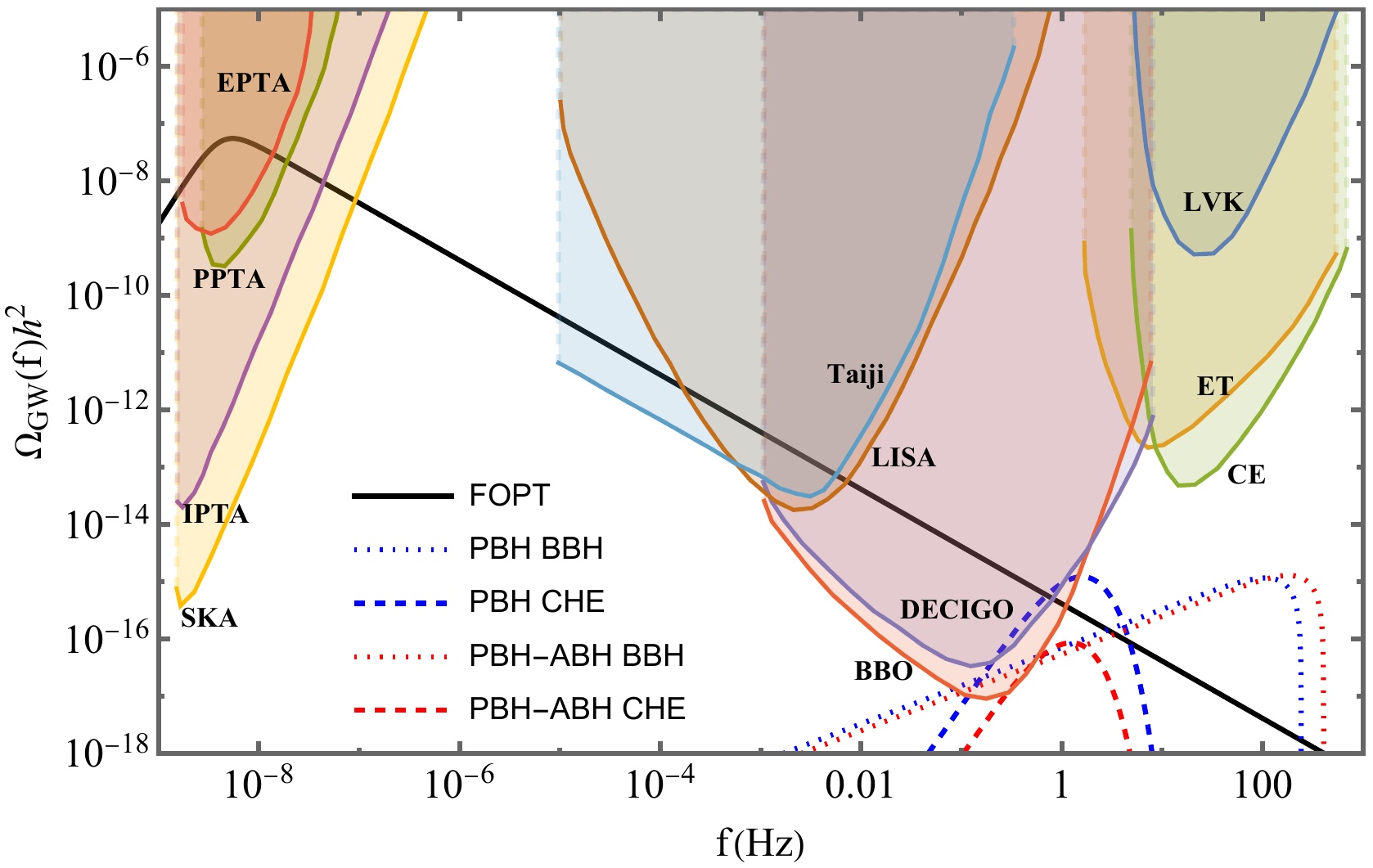}
\caption{The SGWB spectrums originating from different processes for FOPT at $T=60\mathrm{~~MeV}$ which corresponds to the PBHs of mass $40M_{\odot}$.}
\label{cumbkg_1}
\end{figure}

As it can be seen from Fig. \ref{cumbkg_1}, the peak of the spectrum originating from the FOPT is within the reach of the SKA and the PTAs and the tail part of it is within the reach of LISA, Taiji, BBO and DECIGO. The spectrum originating from the PBH-PBH CHE is within the reach of BBO but the spectrum due to the PBH-ABH CHE is not within the reach of any GW detectors. Here, it is also to be noted that the spectrum due to both of these two falls in the same frequency region and due to fact that the amplitude of the spectrum due to the PBH-PBH CHE is much larger than that of the PBH-ABH CHE, the former completely overshadows the later. Therefore, for our subsequent analysis, we will only be considering the spectrum originating due to the PBH-PBH CHE. Similarly, for the case of PBH-PBH and PBH-ABH BBH, it is to be noted that both the spectrum has almost similar shape with respect to the frequencies, but the amplitude of the PBH-PBH BBH SGWB spectrum is slightly higher than that of PBH-ABH BBH in the region where detection is possible, which is why we ignore the spectrum originating from the PBH-ABH binaries in the following subsection as it will not contribute to the net SGWB spectrum in a significant manner. Some part of the PBH-PBH BBH spectrum is within the sensitivity curve of the BBO, but the peak part of it is in the higher frequency region of the spectrum and thus it is inaccessible through BBO and DECIGO. Also, the amplitude is not large enough for it to be detectable through ET and CE. These two issues arise due to the temperature of the FOPT that creates these PBHs as there is a direct dependence of the PBH mass on the FOPT temperature. The peak frequency and the amplitude of the GW spectrum has a direct dependence on the PBH mass. Hence, our focus needs to be shifted to the dependence of these quantities on the transition temperature.
%

\subsection{Transition Temperature and Peak Frequencies}
\label{trans_temp}
\begin{figure}[t]
\centering
\includegraphics[scale=0.45]{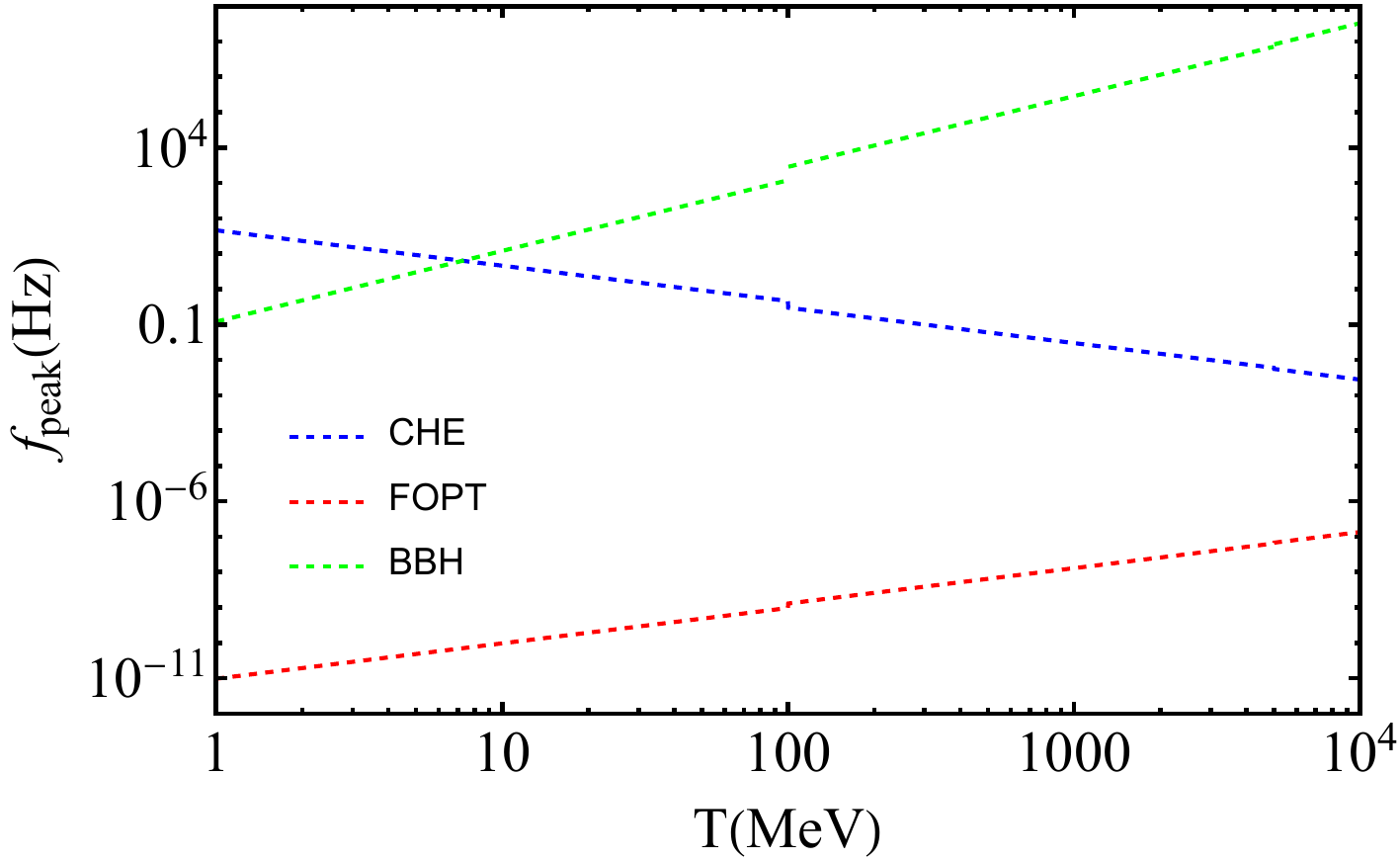}
\caption{The dependences of the peak frequencies of the SGWB spectrum originating from different sources on the temperature of the FOPT.}
\label{peak_freq}
\end{figure}

\begin{figure}[t]
\centering
\includegraphics[scale=0.48]{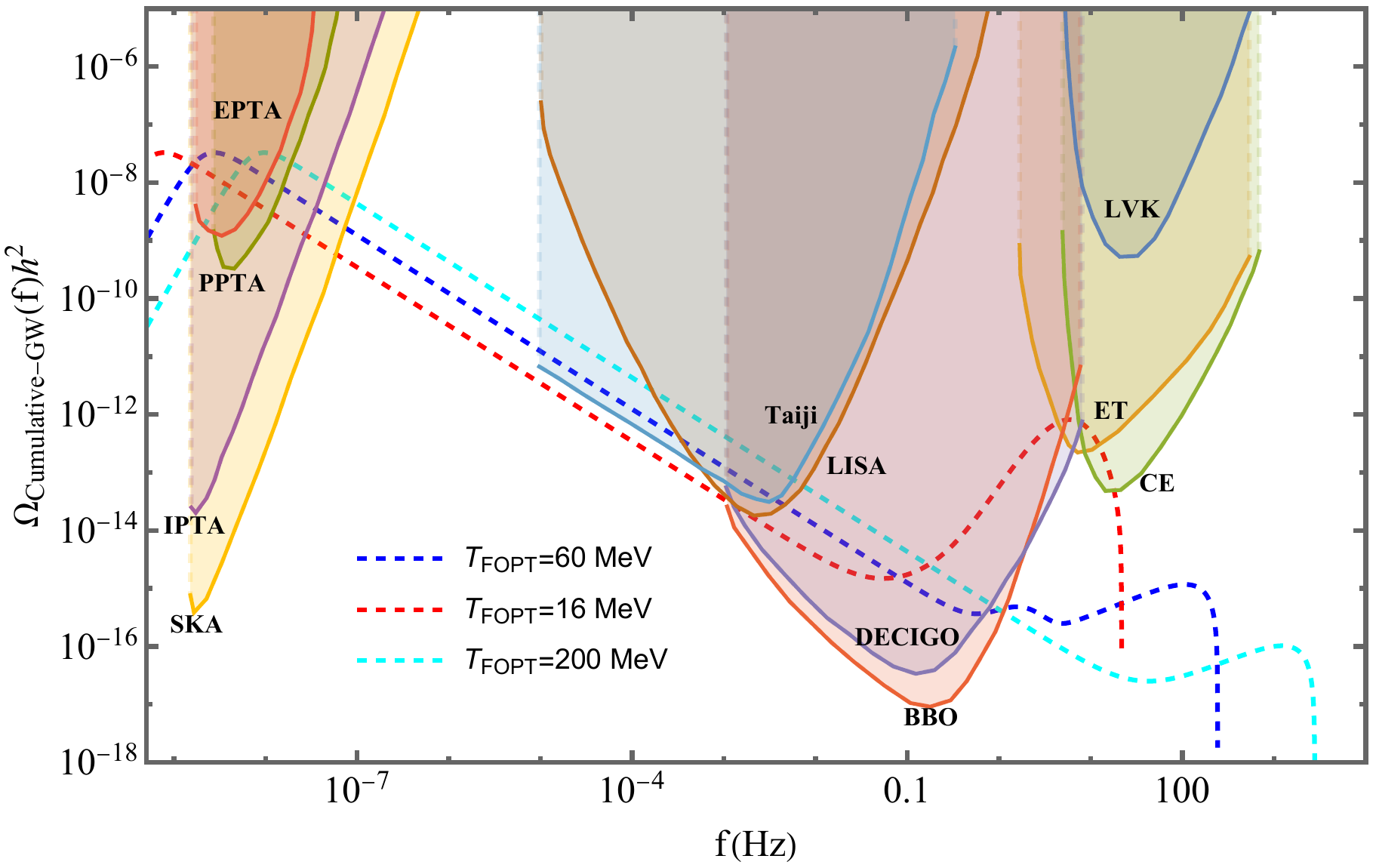}
\caption{The cumulative SGWB spectra originating due to FOPTs at different temperatures.}
\label{cumbkg_diffT}
\end{figure}

The mass of the PBH created during the radiation dominated era due to the collapse of a region of the size of the Hubble horizon can be expressed as~\cite{Carr:2020gox},
\begin{align}
M=1.45\times 10^{5} \left(\dfrac{g_{*}}{10.75}\right)^{-1/2} \left(\dfrac{T}{\mathrm{MeV}}\right)^{-2} M_{\odot},
\end{align}
where $g_{*}$ is the relativistic degrees of freedom at temperature $T$.
Therefore, from Eq.~\eqref{fpeak_bbh} we can write the peak frequency for the SGWB originating from binary PBH systems where the masses of both the participating PBHs are the same as a function of transition temperature,
\begin{align}
f_{\mathrm{peak}}^{\mathrm{BBH}}=0.12 \left(\dfrac{g_{*}}{10.75}\right)^{1/2}\left(\dfrac{T}{\mathrm{MeV}}\right) \mathrm{~Hz}.
\label{fpeak_bbh_T}
\end{align}
Similarly, for the spectrum originating from the CHE of PBHs, using Eq.~\eqref{fpeak_che}, the peak frequency can be expressed as a function of the transition temperature as,
\begin{align}
f_{\mathrm{peak}}^{\mathrm{CHE}}=46  \left(\dfrac{g_{*}}{10.75}\right)^{-1/4}\left(\dfrac{T}{\mathrm{MeV}}\right) \mathrm{~Hz},
\label{fpeak_che_T}
\end{align}
where we have taken the benchmark values for the parameters $a$ and $y$ to be $7\times 10^4\mathrm{~AU}$ and $3.5\times 10^{-5}$ respectively.
Therefore, from Eqs.~\eqref{FOPT_fp}, \eqref{fpeak_bbh_T} and \eqref{fpeak_che_T} we have shown the transition temperature dependence of the peak frequencies in Fig.~\ref{peak_freq}. In generating this Fig. \ref{peak_freq} we have used the benchmark value $\beta/H=3.5$ and we have used $g_{*}=10.75$ for $T<100\mathrm{~MeV}$ and $g_{*}=61.25$ for $100\mathrm{~MeV}<T<1\mathrm{~GeV}$. We can see from this figure that, the peak frequency from the GW spectrum due to the BBH and due to the FOPT increases as the transition temperature increases, whereas for the GW spectrum originating from the CHEs, the peak frequency decreases with increasing transition temperature. It is also to be noted that with the sensitivity reach of different GW detection experiments, signals with frequency ranging from $10^{-9}\mathrm{~Hz}$ to $10^{3}\mathrm{~Hz}$ can be detected. Therefore, from Fig.~\ref{peak_freq} we can see that, if the amplitude of the signal is high enough, then from transition temperature $50\mathrm{~MeV}$ to approximately $10\mathrm{~GeV}$, the peak frequency region of all three signals can be detected, whereas for temperatures lower than that, the peaks of GW signals originating from only BBH and CHE, can be detected. In these cases, even though the peak region of the SGWB spectrum due to the FOPT may not be in the detectable range, most of the tail part will be detectable through the various proposed GW detectors.
So, we shift our focus to the complete cumulative background signals for different transition temperatures in order to understand the detectability of the cumulative SGWB signals from FOPTs. We have shown the cumulative backgrounds for three different FOPTs with transition temperatures $16\mathrm{~MeV}$, $60\mathrm{~MeV}$ and $200\mathrm{~MeV}$ in Fig. \ref{cumbkg_diffT}. In this case, we have taken the benchmark parameters to be $\alpha=1$, $\beta/H=3.5$, $\kappa=0.5$, $a=7\times 10^{4}\mathrm{~AU}$, $y=3.5\times 10^{-5}$ and $f_{\mathrm{PBH}}=0.005$. It is to be noted that, lower the temperature of the FOPT, higher the masses of the PBHs, and therefore higher is the peak in the tail part of the cumulative signal as that consists of the signals from the BBHs and the CHEs. It is also to be seen that the peak of the tail end of the cumulative signal for the FOPT occurring at $T=60\mathrm{~MeV}$ is inaccessible through CE, ET and DECIGO, but that is not the case for the FOPT at $T=16\mathrm{~MeV}$, where the peak of the tail end is within the reach of DECIGO, BBO, ET, and CE. Also, the peak region of the tail end of the cumulative signal due to the FOPT occurring at $T=200\mathrm{~MeV}$ is the lowest among the three, as expected.

It is to be noted that for the cumulative SGWB spectrum shown in Fig. \ref{cumbkg_diffT}, the first peak occurs around $\mathcal{O}(10^{-9}\mathrm{~Hz})$ while the second peak occurs around $\mathcal{O}(1\mathrm{~Hz})$. In the other two cases as well, the peaks are separated by $\mathcal{O}(10^{11}\mathrm{~Hz})$ and $\mathcal{O}(10^{10}\mathrm{~Hz})$ respectively. This very high separation between the peaks makes this signal unique when compared with other exotic SGWB signals with multiple peaks arising from different mechanisms~\cite{Borah:2023iqo,Lozanov:2023aez,Bhaumik:2022pil,Bhaumik:2022zdd}.
%

\section{Summary and Conclusion}
\label{sec:concl}
The central aim of this article is to obtain a prescription to test the claim of creation of PBH during a first order phase transition. In this work, we consider the SGWB from FOPTs and the BBH and CHE of PBHs due to interactions with other PBHs and ABHs. We note that the procedure can easily be extended to other mechanisms which create both PBH and SGWB. 
We have considered two characteristic FOPTs at temperatures $45\mathrm{~GeV}$ and $60\mathrm{~MeV}$ and obtain the mass and taken benchmark abundance of the PBHs which can be expressed as $\{3\times 10^{-5}M_{\odot}$, $0.01\}$ and $\{40M_{\odot}$, $0.005\}$, respectively. Along with this we obtain the SGWB due to these FOPTs. It is to be noted here, that since we are considering non-spinning PBHs with comparatively high masses, gravitational interactions are the only major source through which the secondary SGWB spectrum could arise. This is because of the fact that due to the heavy mass Hawking evaporation does not play any major role and due to non-spinning nature of the PBHs superradiance is immaterial. We leave the discussion of cumulative SGWB due to lighter and spinning PBHs for future work. Then we consider the interaction between PBHs. At first we focus on the SGWB from PBH-PBH binaries for the all the three possible combinations, i.e. two $40M_{\odot}$ PBHs, one $40M_{\odot}$ PBH-one $3\times 10^{-5}M_{\odot}$ PBH and two $3\times 10^{-5}M_{\odot}$ PBHs. From this, we can see that interactions between two $40M_{\odot}$ PBHs creates a SGWB spectrum that can be detected through BBO, whereas for the other two combinations, the spectrum is too weak for even any future GW detectors. We consider the SGWB spectrum due to close hyperbolic interactions between PBHs for the same three combinations of PBH masses. In this case also, we see that for the first PBH mass combination, the resultant SGWB spectrum can be probed through BBO and DECIGO, but the other two combinations can not be probed by any of the proposed detectors. For completeness we also consider the PBH-ABH binaries and PBH-ABH close hyperbolic interactions for both the PBH masses. In this case, we see that SGWB spectrum due to the $40M_{\odot}$ PBH interaction with ABH (both binaries and CHEs) can be probed through BBO, but the signal will be too faint in case of the $3\times 10^{-5}M_{\odot}$ for any of the detectors to detect them. It can also be seen that, the spectrum for the interactions of $40M_{\odot}$ PBHs overshadows the spectrum due to the interactions between the $40M_{\odot}$ PBHs and ABHs. Therefore, we have only considered the results for the interactions between the $40M_{\odot}$ PBHs. It is also to be noted that we have considered benchmark values for CHE in both PBH-PBH and PBH-ABH cases. A detailed analysis regarding the distribution of PBH and ABH will lead to a more realistic and wide parameter space for those variables.
We show the cumulative SGWB due to FOPT and various gravitational interactions of the PBHs. We consider that these PBHs are created due to the FOPT, therefore the SGWB spectrum generated through the interactions of the PBHs are also an indirect consequence of the FOPT itself. Hence, we represent the the sum of all these spectrum as a single cumulative spectrum due to the FOPT. We see that, along with the peak due to the SGWB from the FOPT in the nHz region, secondary peaks also arise around the $1-100\mathrm{~Hz}$ region, some of which could be probed through BBO. A spectrum which spans such a large frequency range and can be detected at different frequencies through different proposed GW detection experiment is unique. Therefore, the detection of a signal of this nature will be a strong indication that FOPTs are indeed the generators of PBH. 
We also investigate the dependence of these different peak frequencies on the temperature at which the FOPT occurs. In this investigation, we found that the peak due to the FOPT increases linearly with the transition temperature, the peak frequency for the binaries is proportional to the square of the transition temperature, whereas the peak frequency due to the CHE is inversely proportional to the temperature. In order to understand the temperature dependence of the cumulative spectrum better, we have presented the cumulative SGWB spectrum due to three different FOPTs occurring at temperatures $16\mathrm{~MeV}$, $60\mathrm{~MeV}$ and $200\mathrm{~MeV}$. We have taken some benchmark values for the other relevant parameters in order to generate the cumulative spectrum. We can see from these three cumulative spectrum, that as the temperature is lower, higher the chances are for the secondary peaks to be in the detection range of the future detectors. In future we would like to study the dark sector FOPTs in details which can lead to FOPTs at temperatures $\mathcal{O}(10\mathrm{~MeV})$. This could give us a more detailed idea regarding some more intricate details of the spectrum and it will also tell us about other consequences of such models. 
In conclusion, we propose that future detection of a spectrum of this kind with multiple detectable peaks spanning a large frequency range will provide crucial information about the creation mechanism of PBH through first order cosmological phase transitions.


%
 
\acknowledgments
IKB thanks Bandita Das for useful discussions. IKB acknowledges the support by the MHRD, Government of India, under the Prime Minister's Research Fellows (PMRF) Scheme, 2022.

\bibliographystyle{JHEP}
\bibliography{PBHGW_ref.bib}

\end{document}